\documentclass[11pt,letterpaper]{article}
\pdfoutput=1
\usepackage{jheppub}

\usepackage{amsmath,slashed}
\usepackage{graphicx}
\usepackage{hepunits}
\usepackage{color}
\DeclareGraphicsExtensions{.pdf,.png,.jpg}

\begin{document} 

\title{Cosmological constraints on Dark Matter models for collider searches}

%\begin{abstract}
%\noindent 
\abstract{
Searches for Dark Matter at the LHC are commonly described in terms of simplified models with scalar, pseudo-scalar, vector and axial-vector mediators.  In this work we explore the constraints imposed on such models from the observed Dark Matter relic abundance.  We present these constraints over a range of mediator masses relevant for the LHC and for future, higher energy colliders.  We additionally compute bounds from a photon line search for the decay of a pseudo-scalar mediator to di-photons that includes the mediator mass region near $750\rm~GeV$.  Finally, we compare cosmological constraints with the reach of a possible future $100~{\TeV}$ circular hadron collider, indirect, and direct detection experiments.
}
%\end{abstract}

\author[b]{Tristan du Pree}
\author[a]{Kristian Hahn}
\author[b]{Philip Harris}
\author[c]{Christos Roskas}
\affiliation[a]{Northwestern University, Evanston IL, USA }
\affiliation[b]{CERN, CH-1211 Geneva 23, Switzerland }
\affiliation[c]{Nikhef, Science Park Amsterdam, Netherlands}
\smallskip
\emailAdd{tristan.dupree@cern.ch}
\emailAdd{kristian.hahn@northwestern.edu}
\emailAdd{philip.coleman.harris@cern.ch}
\emailAdd{croskas@nikhef.nl}

%\pacs{}
%\preprint{XXXX/YY/ZZ}

\maketitle

%%%%%%%%%%%%%%%%%%%%%%%%%%%%%%%%%%%%%%%%%%%%%%%%%%%
\section{Introduction}
\label{sec:intro}
%%%%%%%%%%%%%%%%%%%%%%%%%%%%%%%%%%%%%%%%%%%%%%%%%%%%%

Data collected by the Planck mission~\cite{Planck:2006aa} confirms that dark matter (DM) constitutes nearly 85\% of the total matter content in the universe, corresponding to $\Omega_c\times h^2=0.12$. Under the assumption that both dark and visible matter in the universe are fundamental, DM should be described by a microscopic particle theory\footnote{for a review see e.g.~\cite{Bertone:2004pz}}. The standard model of particle physics (SM) does not contain a viable DM candidate, therefore DM production must be associated with new physics.  The discovery of this physics is among the most important goals in the field. 
%The standard model of particle physics (SM) does not contain any viable DM candidates, and hence the search for the production of DM amounts to one of the most important goals in particle physics. 

%For a large class of models, DM we can reduce the interactions of DM with the SM to a set of well-defined interactions. These interactions go through a mediator, which then connects dark matter to the standard model. 
For a large class of models, DM phenomenology can be reduced to a set
of well-defined DM-SM interactions. These interactions proceed through
a mediator that connects dark matter to the particles of standard
model. At the Large Hadron Collider (LHC), DM searches are performed
using models that describe mediator-based interactions between DM
particles and SM partons.  These interactions can be classified by the
types of messenger fields involved: scalar, pseudo-scalar, vector and
axial-vector. Benchmark models for DM searches at the LHC typically
have mediator couplings to the DM of
$g_{DM}=1$~\cite{Abercrombie:2015wmb,Boveia:2016mrp} and to the SM of
$g_{q}=0.25$ for vector and axial-vector mediators and $g_{q}=1$ for
scalar pseudoscalar
mediators~\cite{Abercrombie:2015wmb,Boveia:2016mrp}.  For the
center-of-mass energies produced at the LHC, sensitivity to the masses
of the mediators and of the DM particles is generally O({\TeV}) for these couplings.  Mediators potentially produced at the LHC are capable of probing a large variety of predicted models, including cosmological predictions of the relic density. 
%\color{red}{KH: The next 2 paragraphs layout what is done for FCC and DD/ID/line-search, so we should add something here for LHC as well. Maybe something like  ``In this work we explore relic density constraints on such models for DM searches at the LHC.'' ? }\color{black}
%.  Searches by the CMS Collaboration at 8~{\TeV}~\cite{CMS-PAS-EXO-12-055} and at 13~{\TeV}~\cite{CMS-PAS-EXO-15-003} employ mediator couplings to the SM of $g_{qq}=1$, whereas the most recent search by the ATLAS Collaboration employs coupling $g_{qq}=0.25$~\cite{ATLAS-CONF-2015-080}.

% not sure about the 2nd use of benchmark ...
%The cosmological constraints on these benchmark models can be treated as a benchmark on the required reach to cover the full phase space of the relic density production. 
Cosmological constraints on these benchmark models can be understood in terms of the reach required to cover the full production phase space of the observed relic density.  The scale of the underlying new physics may be out of reach of the LHC, in which case a hadron collider with higher collision energy is needed.  Many recent studies have explored the physics potential of a future circular collider (FCC) with a center-of-mass energy of 100~{\TeV}~\cite{Arkani-Hamed:2015vfh,Campbell:2013qaa,Hinchliffe:2015qma,Rizzo:2015yha,PhysRevD.47.3065}. We extend our studies of cosmological constraints to DM searches performed with this machine.

We additionally consider the expected bounds of other DM experiments relative to those of the LHC and FCC. The ultimate reach of direct detection experiments is expected to extend to the so-called ``neutrino wall''\cite{Cushman:2013zza,Schumann:2015cpa,Akerib:2015cja,Aalseth:2015mba,Aprile:2013doa,Akerib:2016lao}. We consider the bounds that direct detection experiments with such sensitivity may place on the simplified model framework used in collider searches.  For the case of pseudo-scalar mediators, we consider the impact of projected bounds of indirect detection and photon line searches. 

Our work contributes to the emerging program of DM studies at future colliders in the 100~{\TeV} range~\cite{Low:2014cba,Bramante:2014tba,Xiang:2015lfa,Freitas:2015hsa,Cirelli:2014dsa}. Related studies using simplified models for constraining dark sectors at the LHC include  Refs.~\cite{Buckley:2014fba,Abdallah:2014hon,Malik:2014ggr,Buchmueller:2014yoa,Haisch:2015ioa,Chala:2015ama,Khoze:2015sra,Buchmueller:2015eea,Fan:2015sza,Lebedev:2014bba}.  We also refer readers to the recent summaries~\cite{Abdallah:2015ter,Abercrombie:2015wmb} and references therein.

%%%%%%%%%%%%%%%%%%%%%%%%%%%%%%%%%%%%%%%%%%%%%%%%%%%
\section{Simplified Models}
\label{sec:simpmodel}
%%%%%%%%%%%%%%%%%%%%%%%%%%%%%%%%%%%%%%%%%%%%%%%%%%%%%

DM searches at hadron colliders typically assume that DM particles are pair-produced from the collisions of visible sector particles -- the SM quarks and gluons. In the scenarios studied here there are no direct interactions between the SM sector and the DM particles of the dark sector. Instead, DM-SM interactions are mediated by an intermediate degree of freedom -- the mediator field. In general, four types of mediators (scalar $S$, pseudo-scalar $P$, vector $Z'$ or axial-vector $Z''$) may be involved. The four corresponding classes of simplified models that describe the elementary interactions of these mediators with the SM quarks and DM particles ($\chi$) are 
\begin{align}
\label{eq:LS} 
\mathcal{L}_{\mathrm{scalar}}&\supset\, -\,\frac{1}{2}M_{\rm Med}^2 S^2 - g_{\rm DM}  S \, \bar{\chi}\chi
 - \sum_q g_{SM}^q S \, \bar{q}q  - m_{\rm DM} \bar{\chi}\chi \,,
 \\
 \label{eq:LP} 
\mathcal{L}_{\rm{pseudo-scalar}}&\supset\, -\,\frac{1}{2}M_{\rm Med}^2 P^2 - i g_{\rm DM}  P \, \bar{\chi} \gamma^5\chi
 -\sum_q  i g_{SM}^q  P \, \bar{q}  \gamma^5q  - M_{\rm DM} \bar{\chi}\chi\,,
 \\
 \label{eq:LV} 
\mathcal{L}_{\mathrm{vector}}&\supset \, \frac{1}{2}M_{\rm Med}^2 Z'_{\mu} Z'^{\mu} - g_{\rm DM}Z'_{\mu} \bar{\chi}\gamma^{\mu}\chi -\sum_q g_{SM}^q Z'_{\mu} \bar{q}\gamma^{\mu}q - M_{\rm DM} \bar{\chi}\chi\,,
 \\
 \label{eq:LA} 
\mathcal{L}_{\rm{axial}}&\supset\,  \frac{1}{2}M_{\rm Med}^2 Z''_{\mu} Z''^{\mu} - g_{\rm DM} Z''_{\mu} \bar{\chi}\gamma^{\mu}\gamma^5\chi -\sum_q g_{SM}^q Z''_{\mu} \bar{q}\gamma^{\mu}\gamma^5q - M_{\rm DM} \bar{\chi}\chi\,.
\end{align}
The coupling constant $g_{\rm DM}$ characterizes the interactions of the messengers with the dark sector particles, which for simplicity we take to be Dirac fermions ($\chi$ and $\bar{\chi}$).  The case of scalar DM particles is a straightforward extension of these results. 

The coupling constants linking the messengers to the SM quarks are collectively described by $g^q_{\rm SM}$,
\begin{eqnarray}
{\rm scalar \,\,\& \,\, pseudo-scalar\, messengers:} && \quad g_{\rm SM}^q \equiv\,  g_q\, y_q\,=\, g_q\, \frac{M_q}{v}\,, 
\label{eq:gdef}\\
{\rm vector \,\,\,\& \,\, axial-vector\, \, messengers:} && \quad g_{\rm SM}^q \, =\, g_{\rm SM}\,.
\label{eq:gdef2}
\end{eqnarray}
For scalar and pseudo-scalar mediators, the couplings to quarks are taken to be proportional to the corresponding Higgs Yukawa couplings ($y_q$), as in models with minimal flavour violation~\cite{D'Ambrosio:2002ex}. The $g_q$ scaling factors are assumed to be flavour-universal for all quarks. For vector and axial-vector mediators, $g_{\rm SM}$ is a gauge coupling in the dark sector, which we also take to be flavour universal. The coupling parameters varied are thus $g_{\rm DM}$ and either $g_q$ or $g_{\rm SM}$, depending on the messenger. 

In general, the simplified model description of the dark sector requires five parameters: the mediator mass $M_{\rm Med}$, the mediator width $\Gamma_{\rm Med}$, the dark particle mass $M_{\rm DM}$, and the mediator-SM and the mediator-DM couplings, $g_{\rm SM}$, $g_{\rm DM}$. Our estimate of the mediator width, $\Gamma_{\rm Med}$, uses the assumption that the DM particles and the mediator are the only additions to the SM particle content; this is known as the minimal width assumption.

%%%%%%%%%%%%%%%%%%%%%%%%%%%%%%%%%%%%%%%%%%%%%%%%%%%
\section{Cosmological Constraints for Searches at the LHC}
\label{sec:lhcrelic}
%%%%%%%%%%%%%%%%%%%%%%%%%%%%%%%%%%%%%%%%%%%%%%%%%%%%%

The \verb MadDM ~tool~\cite{Backovic:2013dpa} is used to compute the relic density.  \verb MadDM ~calculates the expected relic DM density in terms of $\Omega_c\times h^2$ for any \verb MadGraph ~model provided~\cite{Alwall:2011uj}.  The tool gives a numerical estimate of the expected relic density based on the standard model of cosmology for any model containing a DM candidate.  The estimate is primarily based on the calculation of the cross-section of the $\chi\chi\rightarrow qq$ process, {\it i.e.} the annihilation of a DM pair into SM particles. For the four mediators explored in this study (generically, $\Phi$), this leads to the annihilation process $\chi\chi\rightarrow\Phi\rightarrow qq$.

\begin{center}
\begin{figure}[h]
\includegraphics[width=0.49\textwidth]{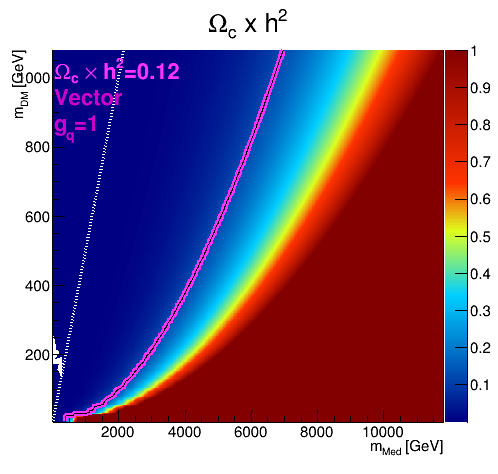} %\hspace{0.5cm} %\hspace{1.5cm}
\includegraphics[width=0.49\textwidth]{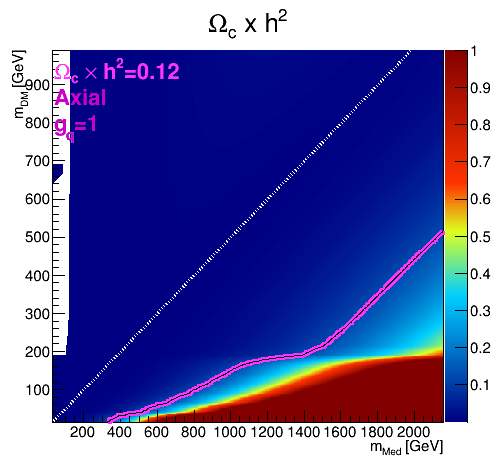} %\hspace{0.5cm} %\hspace{1.5cm}
\includegraphics[width=0.49\textwidth]{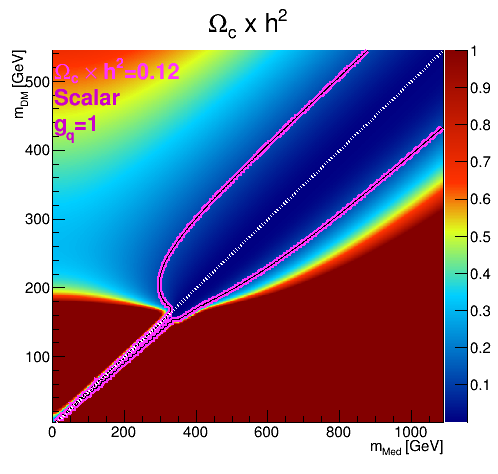} %\hspace{0.5cm} %\hspace{1.5cm}
\includegraphics[width=0.49\textwidth]{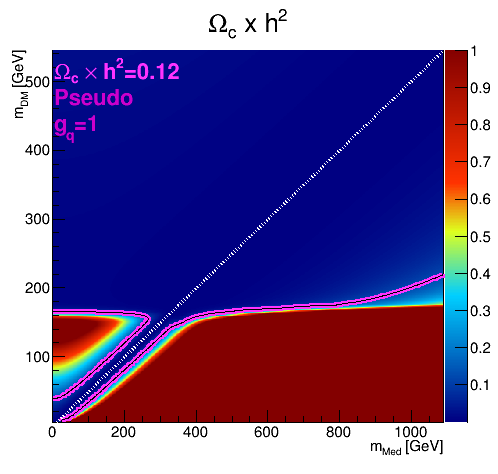} %\hspace{0.5cm} %\hspace{1.5cm}
\caption{The predicted DM relic density for the vector, axial-vector, scalar, and pseudo-scalar mediators for coupling $g_q=g_{SM}=g_{DM}=1$. 
The white dashed line corresponds to the region where $M_{\rm Med}\sim2\times M_{DM}$. 
The pink curve denotes the masses for which the predicted relic DM coincides with the observed $\Omega_c\times h^2=0.12$~\cite{Planck:2006aa}.}
\label{fig:PAS_g1}
\end{figure}
\end{center}

The expected values of $\Omega_c\times h^2$ are shown in Fig.~\ref{fig:PAS_g1} for the mass ranges reachable by the LHC in Run-1 for couplings $g_{DM}=g_{SM}=1$. The expected $\Omega_c$ grows rapidly for $M_\chi<M_t$, which results from a reduced cross section for the $\chi\chi\rightarrow SM$ annihilation process. This feature is particularly strong for models with (pseudo-)scalar mediator due to the enhanced Yukawa couplings to heavy quarks.  The expected $\Omega_c$ also decreases in the region $M_{\rm Med}\sim2\times M_{DM}$, which results from a resonant enhancement in the annihilation cross section.

The pink curves in Fig.~\ref{fig:PAS_g1} correspond to $\Omega_c\times h^2=0.12$, which is the best fit from Planck satellite observations~\cite{plank}. These curves appear (for $g_q=g_{SM}=1$) in searches by the CMS Collaboration~\cite{CMS-PAS-EXO-12-055,CMS-PAS-EXO-15-003}.  The regions closer to the line $M_{\rm Med}\sim2\times M_{DM}$ have lower values of $\Omega_c\times h^2$ and correspond hence to under-abundant DM production. The regions away from the line $M_{\rm Med}\sim2\times M_{DM}$ have higher values of $\Omega_c\times h^2$ and correspond to DM overabundance.

\begin{center}
\begin{figure}[h]
\includegraphics[width=0.49\textwidth]{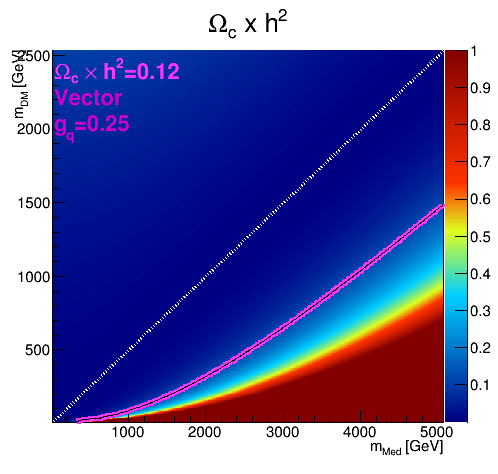} %\hspace{0.5cm} %\hspace{1.5cm}
\includegraphics[width=0.49\textwidth]{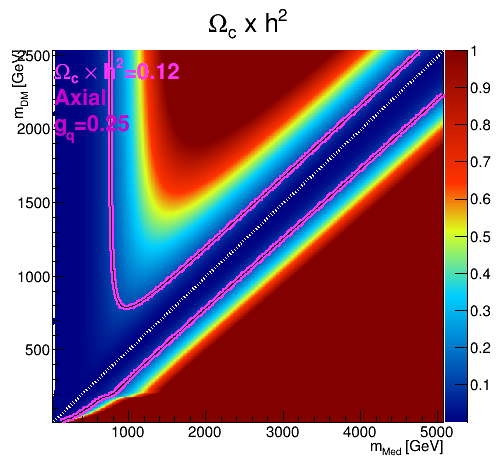}  %\hspace{0.5cm} %\hspace{1.5cm}
\caption{The predicted DM relic density for the vector and axial-vector mediators for coupling $g_{SM}=0.25$ (labelled $g_{q}=0.25$) and $g_{DM}=1$.  
The white dashed line corresponds to the region where $M_{\rm Med}\sim2\times M_{DM}$. 
The pink curve denotes the masses for which the predicted relic DM coincides with the observed $\Omega_c\times h^2=0.12$.
}
\label{fig:PAS_g25}
\end{figure}
\end{center}

Fig.~\ref{fig:PAS_g25} shows the predicted values of $\Omega_c$ for $g_{SM}=0.25$ (labelled $g_{q}=0.25$). These coupling values are currently recommended by the LHC DM WG~\cite{Abercrombie:2015wmb,Boveia:2016mrp} and are used in a recent 13~{\TeV} DM search by the ATLAS Collaboration~\cite{ATLAS-CONF-2015-080}. The behavior of $g_{SM}=0.25$ results are similar to those of $g_{SM}=1$:  annihilation is enhanced for $M_{\rm Med}\sim2\times M_{DM}$ and suppressed for $M_{DM}<M_t$.  The expected relic density is smaller than that for $g=g_q=g_{SM}=g_{DM}=1$ due to the decrease in annihilation cross section. 

Compared to the constraints for the (axial-)vector mediators, the constraints for the (pseudo-)scalar mediators are, for low mass, closer to the line $M_{\rm Med}\sim2\times M_{DM}$. This is attributed to the relatively narrow width of the (pseudo-)scalar mediators, and the Yukawa nature of its couplings to the SM particles.
The behavior at $M_{\rm Med}\sim800~{\GeV}$ for the axial mediator is due to double-mediator production, which occurs when $M_{DM}\geq M_{\rm Med}$~\cite{Kahlhoefer:2015bea}.

The results have also been cross-checked against an independent analytical estimate of the relic density and the results were found in agreement~\cite{Haisch:2015ioa}. 

%\clearpage

%%%%%%%%%%%%%%%%%%%%%%%%%%%%%%%%%%%%%%%%%%%%%%%%%%%
\section{Cosmological Constraints for a 100 TeV Collider}
\label{sec:fccrelic}
%%%%%%%%%%%%%%%%%%%%%%%%%%%%%%%%%%%%%%%%%%%%%%%%%%%%%

%KH comment: If the scale of new physics is out of range of the LHC's reach, a (hadron) collider with higher collision energy is needed~\cite{Arkani-Hamed:2015vfh}. 
A (hadron) collider with higher collision energy and mass reach would be needed if the scale of the new physics underlying DM production lies beyond the reach of the LHC~\cite{Arkani-Hamed:2015vfh}. At such collision energy, the sensitivity to $M_{\rm Med}$ typically 
extends up to few~{\TeV} (for the scalar and pseudo-scalar types) or $>15~\TeV$ (for the vector and axial-vector types)~\cite{Harris:2014hga}.

\begin{center}
\begin{figure}[h]
\includegraphics[width=0.49\textwidth]{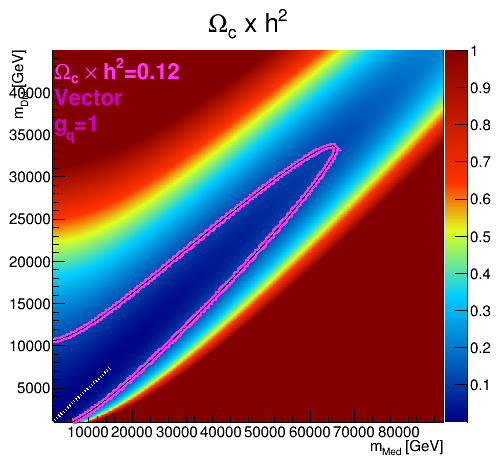}
\includegraphics[width=0.49\textwidth]{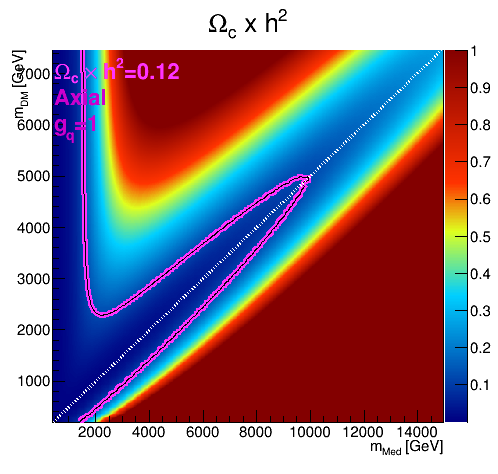}
\includegraphics[width=0.49\textwidth]{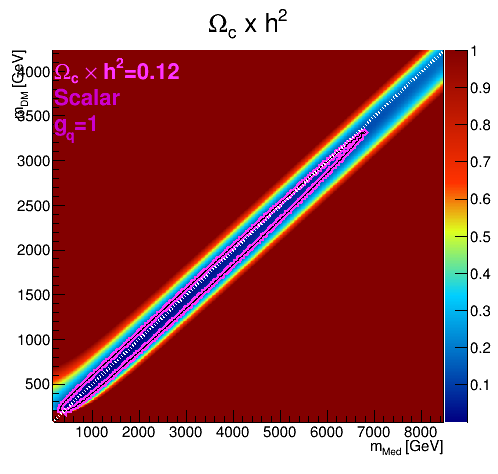}
\includegraphics[width=0.49\textwidth]{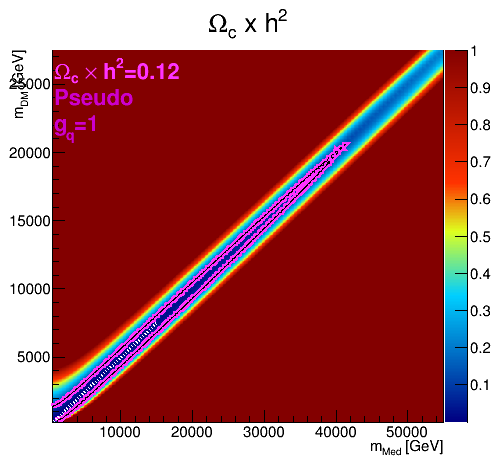}
\caption{Predicted DM relic density for the vector, axial-vector, scalar and pseudo-scalar mediators for default coupling $g_q=g_{SM}=g_{DM}=1$. 
The white dashed line corresponds to the region where $M_{\rm Med}\sim2\times M_{DM}$. 
The pink curves denote the masses for which $\Omega_c\times h^2=0.12$.}
\label{fig:FCC_g1}
\end{figure}
\end{center}

The predicted relic DM density is shown in Fig.~\ref{fig:FCC_g1} for $g=1$ over a wide mass range characteristic of the FCC.  The region with small predicted values of $\Omega_c\times h^2$ lie mostly around the diagonal $M_{\rm Med}\sim2\times M_{DM}$. This is because resonant annihilation of $\chi\chi\rightarrow SM$ is preferred for $M_{\rm Med}\sim2\times M_{DM}$.  Moreover, the constraint tends to align more closely to the diagonal for (pseudo-)scalar mediators than for (axial-)vector mediators, due to the narrower widths of the former.  
%The behavior at $M_{\rm Med}\sim800~{\GeV}$ for the axial mediator is due to double-mediator production, which occurs when $M_{DM}\geq M_{\rm Med}$~\cite{Kahlhoefer:2015bea}.

\begin{center}
\begin{figure}[h]
\includegraphics[width=0.49\textwidth]{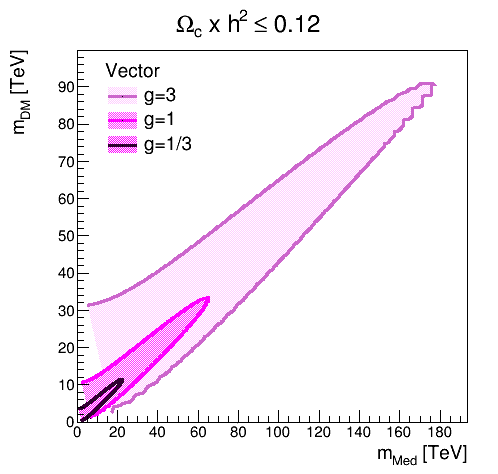} %\hspace{0.5cm} %\hspace{1.5cm}
\includegraphics[width=0.49\textwidth]{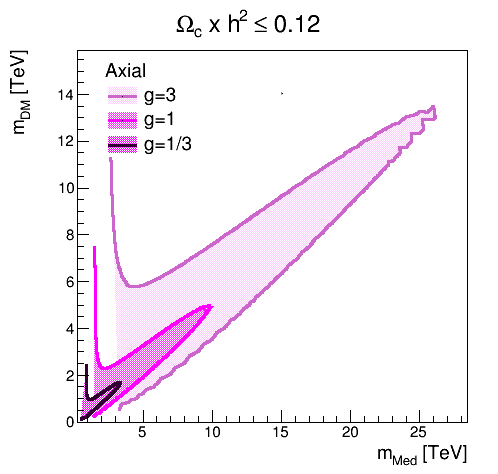}  %\hspace{0.5cm} %\hspace{1.5cm}
\includegraphics[width=0.49\textwidth]{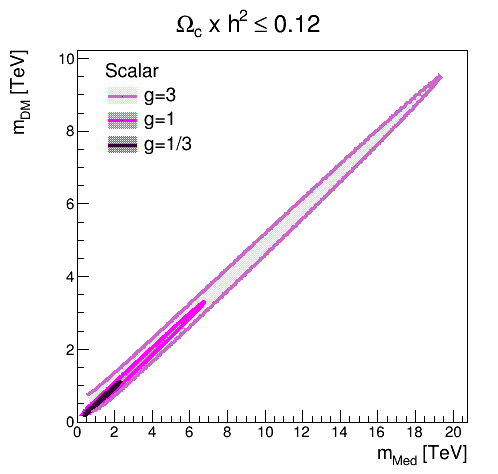} %\hspace{0.5cm} %\hspace{1.5cm}
\includegraphics[width=0.49\textwidth]{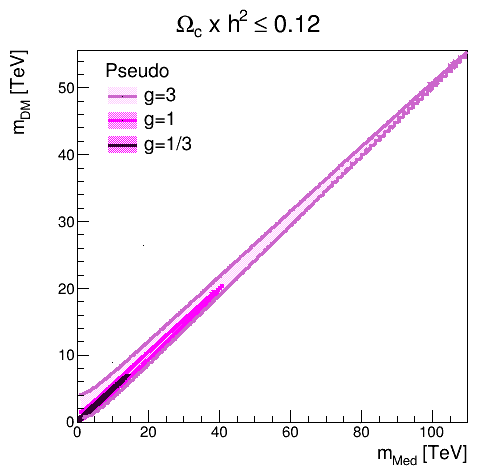} %\hspace{0.5cm} %\hspace{1.5cm}
\caption{Predicted regions for $\Omega_c\times h^2\leq0.12$ for the vector, axial-vector, scalar and pseudo-scalar mediators for various couplings: $g_{DM}=1/3,1,3$. The label $g$ on the plots denotes $g_{DM}$, for all cases $g_q=g_{SM}=1$.}
\label{fig:FCC_gvar}
\end{figure}
\end{center}

In general, the results for these models indicate a cosmologically preferred region of masses up to $M_{\rm Med}<7-10~${\TeV} (for scalar and axial-vector mediators) and $M_{\rm Med}<40-65~${\TeV} (for vector and pseudo-scalar mediators). The bounds from $\Omega_c\times h^2\leq0.12$ are considered as a function of the couplings in Fig.~\ref{fig:FCC_gvar}. Although the shapes of the constraints do not change significantly for the couplings considered, the maximally allowed mass changes significantly, scaling in proportion with the coupling.

\clearpage
%%%%%%%%%%%%%%%%%%%%%%%%%%%%%%%%%%%%%%%%%%%%%%%%%%%
\section{Direct Photon search}
\label{sec:directphoton}
%%%%%%%%%%%%%%%%%%%%%%%%%%%%%%%%%%%%%%%%%%%%%%%%%%%%%

\begin{center}
\begin{figure}[h]
\includegraphics[width=0.49\textwidth]{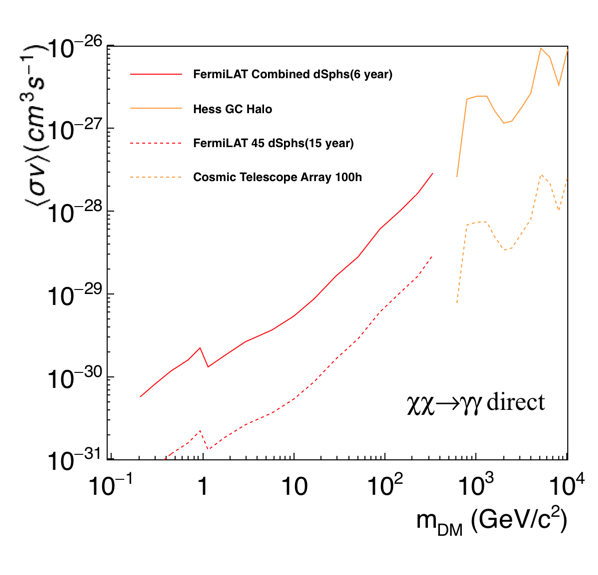}  %\hspace{0.5cm} %\hspace{1.5cm}
\includegraphics[width=0.49\textwidth]{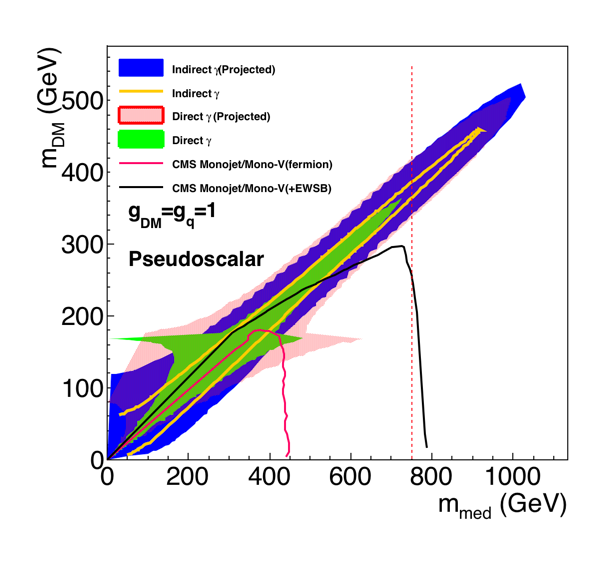} %\hspace{0.5cm} %\hspace{1.5cm}
\caption{Left: Photon line bounds from FermiLAT~\cite{Drlica-Wagner:2015xua} and
  HESS~\cite{HESS:2015cda} - the dashed lines show the expected updated
  sensitivity based on 15 years of running for
  FermiLAT~\cite{Drlica-Wagner:2015xua} and the Cosmic Telescope
  Array~\cite{cta}. The excluded regions are above the curves, \textit{i.e.} the regions with largest cross-section
  Right: the translation of the FermiLAT, HESS and
  photon line bounds to the ($M_{\rm Med},M_{\rm DM}$) plane. The yellow line
  and blue shaded regions correspond to the indirect photon searches
  and the red and green shaded regions correspond to the  photon
  line searches. The collider bounds from the most sensitive CMS DM search, the combined mono-jet, mono hadronic vector boson search, are
also shown~\cite{CMS-PAS-EXO-12-055}. This is presented for both the simplified model where a pseudoscalar coupling to fermions is present (red), and the case where pseudoscalar vector boson couplings are also present(black). For the latter, additional physical effects from the extended model are neglected. A red line at 750~{\GeV} is added
to guide the eye.
 }
\label{fig:ggBound}
\end{figure}
\end{center}

The scattering cross section of a pseudo-scalar mediator is heavily velocity-suppressed at direct detection experiments.  Because of this, we compare collider and relic constraints with bounds from indirect detection. We consider indirect detection bounds from searches for photons resulting both from the decay of SM particles produced in DM annihilation, and from those from direct mediator-to-photon production, {\it i.e.}  ``photon line'' searches. Bounds from the latter are computed by considering the direct production of a pseudo-scalar mediator to photons through a top loop. The velocity averaged annihilation cross section to photons,$\langle\sigma v\rangle_{\gamma}$, can be expressed as ~\cite{Buckley:2014fba,Low:2015qep}:

\begin{eqnarray}
\langle\sigma v\rangle_{\gamma} & = &
                                      \frac{1}{4\pi}\left(\frac{\alpha}{2\pi}\right)^{2}\frac{g_{q}^{2} y_{t}^{2}}{v^{2}}\frac{g^{2}_{DM}}{(M^{2}_{\rm Med}-4m^{2}_{DM})^{2}+M_{\rm Med}^{2}\Gamma^{2}_{\rm Med}} \left|N_{c}Q_{t}^{2}F_{A}\left(\frac{m^{2}_{t}}{m^{2}_{DM}}\right)\right|^{2},{\rm\,} \\
F_{A}(\tau) & = & \tau f(\tau), {\rm \, and }  \\
f(\tau)      & = & \theta(\tau-1)\arcsin^{2}\left(\frac{1}{\sqrt{\tau}}\right)-\theta(1-\tau)\frac{1}{4}\left(\log\frac{1+\sqrt{1-\tau}}{1-\sqrt{1-\tau}}-i\pi\right)^{2}.
\end{eqnarray}
Here, 
%$M_{\rm Med}$ is the mass of the mediator, $M_{DM}$ is the
%mass of the DM, $g_{DM}$ is the DM coupling,
%$\Gamma_{med}$ is the width of the mediator, 
$N_{c}$ is the number of colors, $Q_{t}$ is the top charge, $g_{q}$ is the coupling to quarks,
%$m_{t}$ is the top mass, 
and $y_{t}/v$ is the Yukawa coupling divided
by the Higgs vacuum expectation. From this cross section formula, we can directly compare with the photon line searches\cite{Fan:2013faa,Cohen:2013ama} from HESS~\cite{Abramowski:2013ax} and FermiLAT~\cite{Ackermann:2015lka,Buckley:2014fba} in Fig.~\ref{fig:ggBound}~(left). 

The results in Fig.~\ref{fig:ggBound}~(right) compare the bound from the photon line search with bounds from indirect searches.  The photon line bound is less sensitive than, although comparable to, the indirect bounds.  In addition, we observe that the current photon line bound approaches sensitivity to a 750~{\GeV} mediator mass for $g_{DM}=g_{q}=1$. This is close to the expected sensitivity of the excess of diphoton excess observed at the
LHC~\cite{ATLAS-CONF-2016-018,CMS-PAS-EXO-16-018,Backovic:2015fnp,Mambrini:2015wyu,DiChiara:2016dez,Redi:2016kip,Ge:2016xcq,DeRomeri:2016xpb,Bharucha:2016jyr,Salvio:2016hnf,Hektor:2016uth,D'Eramo:2016mgv,Deppisch:2016scs,Palle:2015vch,Ghorbani:2016jdq,Huang:2015svl,Moretti:2015pbj,Dev:2015isx,Park:2015ysf,Low:2015qep,Franceschini:2015kwy,Ellis:2015oso,Bai:2015nbs}. For a DM coupling of order unity, the potential reach of direct photon searches may provide for a detection of pseudo-scalar mediated DM in the near future. 

We additionally consider the projected sensitivities of FermiLAT~\cite{ChangJin:550,Dokuchaev:2015ghx,Ackermann:2015lka} and of the upgraded HESS experiment~\cite{Abramowski:2013ax}. We plot the direct and indirect bounds for these projected results in Fig.~\ref{fig:ggBound}.  From these projected bounds, one observes an extension in the sensitivity to a mediator mass of 1~{\TeV}.  We compare these results with relic density bounds in next Section.

%%%%%%%%%%%%%%%%%%%%%%%%%%%%%%%%%%%%%%%%%%%%%%%%%%%
\section{Experiments vs Relic constraints}
\label{sec:relicvsfcc}
%%%%%%%%%%%%%%%%%%%%%%%%%%%%%%%%%%%%%%%%%%%%%%%%%%%%%

\begin{center}
\begin{figure}[h]
\includegraphics[width=0.49\textwidth]{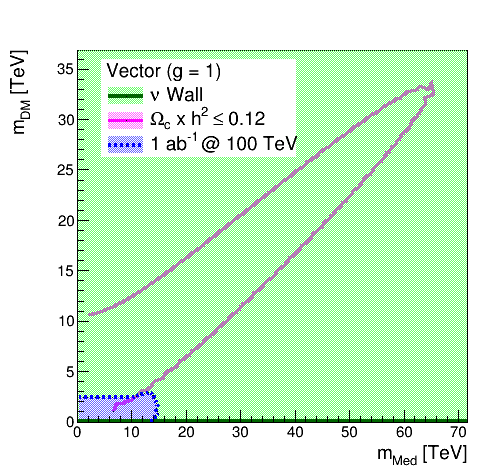}
\includegraphics[width=0.49\textwidth]{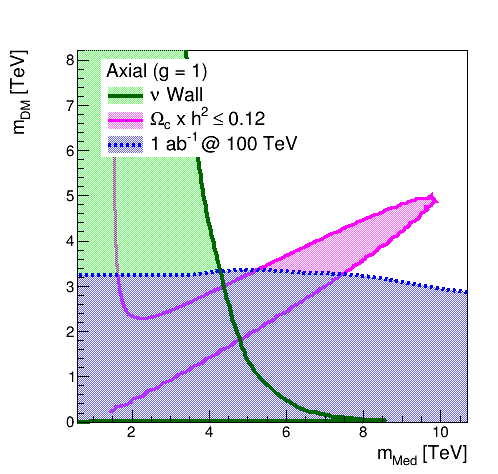} 
\includegraphics[width=0.49\textwidth]{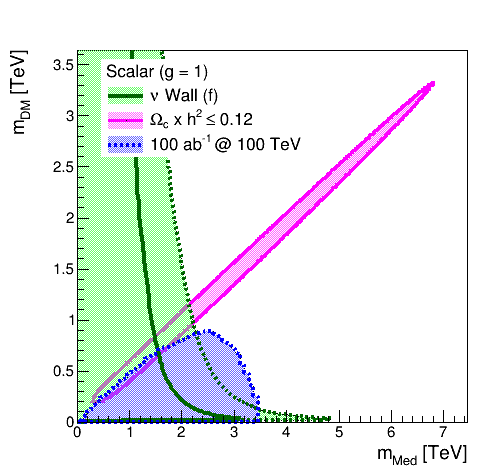}
\includegraphics[width=0.49\textwidth]{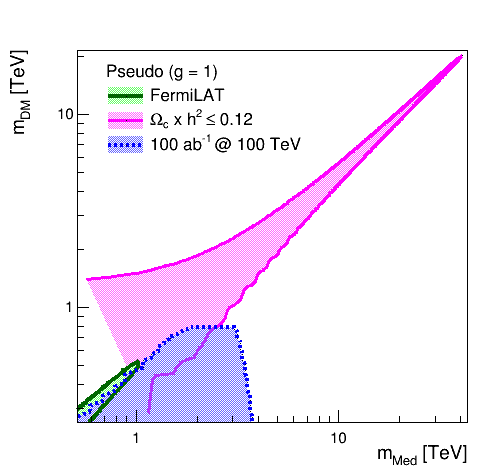}
\caption{Pink: regions with $\Omega_c\times h^2\leq0.12$ for the vector, axial-vector, scalar and pseudo-scalar mediators, for coupling $g_q=g_{SM}=1$, and $g_{DM}=1$.
Green: neutrino wall. Blue: expected sensitivity for a 100~{\TeV} collider with 1~ab$^{-1}$~\cite{Harris:2015kda}.}
\label{fig:moneyplot}
\end{figure}
\end{center}

Fig.~\ref{fig:moneyplot} compares cosmological constraints and bounds from direct detection with the 100~{\TeV} collider reach for 1-100~ab$^{-1}$ of data. The reach of the FCC (in blue) extends to $M_{Med}\sim35~${\TeV} for a scalar mediator and up to $M_{Med}\sim~$15~{\TeV} for vector and axial-vector mediators.  The collider constraint for a pseudo-scalar mediator is less stringent, reaching only to $M_{Med}\sim4~${\TeV}~\cite{Harris:2015kda}.

Projected FCC constraints do not completely cover the cosmologically allowed region of DM parameter space.  Nevertheless, the axial-vector model is almost fully accessible at the FCC, particularly if the large datasets expected for such an experiment are ultimately obtained.  A significant fraction of parameter space can also be probed for scalar-mediated models. Pseudo-scalar mediators pose the most significant challenge; as Fig.~\ref{fig:moneyplot} shows, both FCC and indirect detection experiments are incapable of constraining the parameter space allowed by relic density observations.

The sensitivity of a 100~{\TeV} collider decreases for smaller coupling values,  however cosmological constraints are impacted more significantly.  This results in tighter collider constraints for all mediators. Direct searches for mediator decays to standard model particles can provide even tighter constraints.  Examples of such searches include those for axial mediator decays to dijets and scalar mediator decays to di-photons.  Overall, the pseudo-scalar mediator is perhaps the most challenging model to cover with experimental searches - the strongest handles come from collider based searches.

\clearpage

%%%%%%%%%%%%%%%%%%%%%%%%%%%%%%%%%%%%%%%%%%%%%%%%%%%
\section{Conclusions}
\label{sec:conclusions}
%%%%%%%%%%%%%%%%%%%%%%%%%%%%%%%%%%%%%%%%%%%%%%%%%%%%%

Cosmological constraints have been shown for the class of simplified mediator models used in searches at the LHC. The numerical predictions of relic DM abundance are calculated using \verb MadDM ~, which uses the \verb MadGraph ~simplified models to estimate the cross-section of the $\chi\chi\rightarrow SM$ annihilation within the standard model of cosmology.

For DM and mediator masses accessible at the LHC, the shapes of the relic constraints are attributed to suppressed mediator to dark matter decays when $M_{DM}<M_t$ and enhanced resonant annihilation for $M_{Med}\sim2\times M_{DM}$. Cosmological constraints have also been shown for the wider mass ranges reachable at possible future colliders.  Masses consistent with cosmological observations typically reach up to 10-100~{\TeV}.  The shapes of the bounds in this wide mass range are attributed to resonant annihilation and the width of the mediators.

The LHC has sensitivity to a small part of the cosmologically preferred parameter space for the models considered. A 100~{\TeV} collider, on the other hand, has significant sensitivity in the full parameter space allowed by relic density constraints.  Significant coverage is obtained for scalar and axial-vector mediators, whereas the pseudo-scalar mediated model is rather difficult to constrain. 

The cosmological constraints presented in this document are available for the collider searches by the LHC Collaborations at~\cite{relicWebpage}.

\section*{Acknowledgments}
\noindent 
We would like to thank Mihailo Backovic for his help in running \verb MadDM. ~We would like to thank Jessie Shelton for motivating the direct photon search. 
We would like to thank Michelangelo Mangano, Filip Moortgat, and Albert de Roeck for valuable discussions.
TdP would like to thank Kevin Sung for feedback on the mediator models and Nicholas Wardle for visualization suggestions.
PH would like to thank Valentin Khoze, Ciaran Williams, and Michael Spannowsky for the initial work on higher-energy colliders that led to this paper. 

\bibliographystyle{jhep}
\bibliography{RelicVsFCC_bib}
\end{document}